\documentclass{optica-article}

\journal{opticajournal} 

\articletype{Research Article}

\usepackage{graphicx}
\usepackage{dcolumn}
\usepackage{bm,soul}
\usepackage{units}
\usepackage{qcircuit}

\newcommand{\ket}[1]{\ensuremath{\vert{#1\rangle}}} 
\newcommand{\bra}[1]{\ensuremath{{\langle #1}\vert}}

\newcommand{\ketbra}[2]{\ensuremath{|{#1 \rangle}{\langle #2}|}}
\newcommand{\op}[1]{\hat{#1}}

\newcommand{\I}{\text{i}}
\newcommand{\E}{\text{e}}

\begin{document}
\pagenumbering{arabic} 

\title{Active polarization stabilization of fields in an optical fiber for protective measurements}

\author{E. Pascoe, A. Catalan, J. Sharkansky and M. Beck\authormark{*} }

\address{Department of Physics, Reed College, 3203 SE Woodstock Boulevard, Portland, OR 97202, USA}

\email{\authormark{*}beckm@reed.edu} 


\begin{abstract*} 
We have performed Zeno protective measurements of quantum polarization states by coupling the polarization to a temporal pointer (arrival time) in a birefringent optical fiber. It is necessary to actively stabilize the polarization, and we do this by using the signal photon counts themselves as the error signal in a feedback loop. We compare these measurements to a stabilization scheme using a classical reference beam as the error signal. The method using photon counts has higher signal levels and significantly reduced background. These improvements allow us to increase the number of Zeno stages in our measurements from 9 to 13, with a corresponding decrease in the measurement uncertainty.

\end{abstract*}

\section{Introduction}
Weak measurements are a form of quantum measurement that involve both a preselection (state preparation) and postselection procedure, with a weak interaction between the signal and an apparatus in the middle \cite{Aharonov_1988}. Since their introduction, weak measurements have seen considerable theoretical and experimental attention \cite{Brunner_2003, Brunner_2004, Wang_2006, Hosten_2008, Dixon_2009a, Tamir_2013,Dressel_2014,Rebufello_2025}. 

A protective measurement (PM) is a type of weak measurement in which the quantum state does not change appreciably during the measurement \cite{Aharonov_1993a, Aharonov_1993, Haridass_1999, Piacentini_2017, Rebufello_2021a, Chen_2023, Schlosshauer_2024}. A PM on a single system yields the expectation value of an observable, so it is possible to obtain a determination of the expectation value using only a single measurement, rather than an ensemble \cite{Piacentini_2017}. Protective measurements can also be used to determine the eigenstates of an otherwise unknown Hamiltonian \cite{diosi_2014}.

There has been some controversy in the literature about protective measurements, regarding what they imply about the meaning of the wavefunction and/or what $a\: priori$ knowledge is needed about the quantum state \cite{Unruh_1994, Rovelli_1994, Alter_1995, Dariano_1996}. In our experiments we use an active protection scheme based on the quantum Zeno effect \cite{Misra_1977, Itano_1990, Kwiat_1999b, Itano_2019}, which employs a series of “Zeno stages” consisting of a weak interaction and a projective measurement. In this scheme the state of the system does not change at all, since it is repeatedly projected back onto the same state. We have argued previously that we have no $a\: priori$ knowledge of the state in our measurements \cite{Chen_2023}. However, in our Zeno protection scheme it is necessary that the preselected and postselected states be the same, and this is accomplished in our experiments by performing projective measurements for each. As such, we do know that our state was produced by a projective measurement that can be repeated many times.

Zeno protective measurements of quantum polarization states have been performed by weakly coupling the polarization to either spatial \cite{Piacentini_2017, Rebufello_2021a} or temporal \cite{Chen_2023} degrees of freedom. In our previous experiments we coupled polarization states to arrival times by using a birefringent optical fiber to provide a differential group delay (DGD) between $\ket{H}$ and $\ket{V}$ states \cite{Chen_2023}. Multiple Zeno stages were implemented by sending photons around a fiber loop, with each loop corresponding to a single Zeno stage. In these experiments it was necessary to actively stabilize the polarization in order to circumvent polarization drifts due to environmentally-induced phase shifts in the fiber. The stabilization also ensured that the postselection measurement corresponded to the same initial state that was produced by the state preparation. Experiments using a spatial pointer involve only free-space optics, and do not require active polarization stabilization, but have so far been limited to fewer Zeno stages than experiments using a temporal pointer \cite{Piacentini_2017, Rebufello_2021a}.

There have been many previous experiments that have demonstrated active polarization stabilization for transmitting quantum states over optical fibers. Most of these have been performed with quantum-key distribution or quantum networking applications in mind. Some experiments use a classical reference signal to provide the error signal that is used to stabilize the polarization; this signal can be separated from the quantum signal using time, wavelength or propagation-direction multiplexing \cite{Xavier_2008, Chen_2023, Craddock_2024, Wang_2024, Gul_2025}. Other experiments use the quantum signal itself as the error signal. Most often these techniques use what we refer to as an "interruption" protocol, which is essentially a form of time-division multiplexing. Here the quantum signal transmission is occasionally interrupted by using the signal photons to provide information about the polarization state instead \cite{Peng_2007, Chen_2007, Ding_2017, Agnesi_2020, Neumann_2022, Mayboroda_2024, Mantey_2025}.

In our previous PM experiments we stabilized the polarization using a classical reference beam that counter-propagated with the signal beam \cite{Chen_2023}. However, this reference beam introduced background photons that limited the performance of our measurements. Perhaps a different scheme for multiplexing the signal and reference could reduce this background, but it can never be completely eliminated. Furthermore, multiplexing a classical reference and a quantum signal will always require additional hardware that will invariably introduce loss. In Ref. \cite{Chen_2023} we used two circulators for this purpose, and the loss that they introduced further limited our measurement performance.

For these reasons we have implemented a scheme that uses the detected single-photon-level signal itself to provide the error signal for active polarization stabilization. However, our technique differs from the time-multiplexed schemes described above in that we do not need to interrupt the signal. The same photons that are used for our measurements are simultaneously used to stabilize the polarization. This new scheme reduces loss, and nearly eliminates background. These improvements allow us to increase the number of Zeno stages in our PM from 9 to 13, correspondingly reducing the uncertainty in our measurements. Thirteen is the largest number of Zeno stages in any PM experiment to date. The dramatically reduced background allows us to make these improvements while also significantly reducing the averaging time in our experiments.  

This paper is organized as follows. In Section~\ref{sec:theory} we describe the theory associated with both our Zeno PM scheme and our polarization stabilization technique. In Section~\ref{sec:expt} we present our experimental results. First we show that our stabilization technique does indeed maintain the polarization state of our field with high fidelity. Next we demonstrate that we are able to successfully perform PMs with 13 Zeno stages. In section~\ref{sec:conclusion} we provide some concluding remarks. 

\section{\label{sec:theory}Theory}

\subsection{\label{sec:PMtheory}Protective measurements}

\begin{figure*}[ht!]
\begin{equation*}
\centering
\Qcircuit @C=1em @R=.7em {
\lstick{\mathcal{S}: \ket{\psi_0}}     & \multigate{1}{\op{U}} &  \measure{\ket{\psi_0}} & \qw& \multigate{1}{\op{U}} &  \measure{\ket{\psi_0}} & \qw & \push{\cdots\quad}  & \multigate{1}{\op{U}} &  \measure{\ket{\psi_0}} & \qw \\
\lstick{\mathcal{A}}     & \ghost{U}&  \qw &\qw& \ghost{U}&  \qw & \qw & \push{\cdots\quad} & \ghost{U}&  \qw & \meterB{\langle\op{O}\rangle}
\gategroup{1}{2}{2}{3}{.7em}{--} 
\gategroup{1}{5}{2}{6}{.7em}{--} 
\gategroup{1}{9}{2}{10}{.7em}{--} 
}
\end{equation*} 
\caption{\label{fig:Zeno}Principle of a Zeno protective measurement. After each weak interaction with the measuring device, the system is projected back onto its initial state $\ket{\psi_0}$. Each dashed unit represents a Zeno stage.}
\end{figure*}
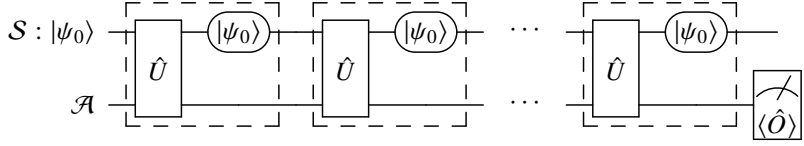

Figure~\ref{fig:Zeno} illustrates the basic concept of a Zeno protective measurement. A quantum system $\mathcal{S}$ is prepared in an initial state $\ket{\psi_0}$. A unitary operator $\op{U}$ weakly couples this state to a classical apparatus $\mathcal{A}$. A measurement that projects the quantum system back onto $\ket{\psi_0}$ is then performed, protecting the state. The action of weak coupling and projection is repeated many times. At the end the apparatus pointer is read, yielding the expectation value of an observable of the system $\langle\op{O}\rangle$.

A detailed theory describing our experiments is presented in Ref.~\cite{Chen_2023}, and we present a briefer version here. We consider a field whose initial temporal wavepacket state $\ket{\phi(0)}$ is assumed to be a Gaussian of duration $\tau_{G}$, and whose polarization state is $\ket{\psi_0}  = \cos\theta\ket{H}+\E^{\I\varphi} \sin\theta\ket{V}$. This wavepacket passes through a birefringent material of length $L$. The $H$ and $V$ components of the wavepacket travel at different group velocities, so after exiting at $L$ they have separated by an amount $\tau$. Define the average travel time $t_\text{ave}$ to be the travel time through the medium at the average group velocity. Assume that the $V$ component propagates faster, so it arrives at $L$ at $t_V=t_\text{ave}-\tau/2$ and the $H$ component arrives at $t_H=t_\text{ave}+\tau/2$. Now we normalize all times to the pulse duration: $\tilde{t}=t/\tau_G$, and $\tilde{\tau} =\tau/\tau_G$.

It was shown in Ref.~\cite{Chen_2023} that we can express the interaction between the quantum system and the apparatus in terms of a unitary operator $\op{U}(\tilde{\tau}) = \exp\left[ -\I \frac{\tilde{\tau}}{2} \op{O} \otimes \op{A} \right]$. Here $\op{O} = \ketbra{H}{H} - \ketbra{V}{V}$ is a polarization observable, $\op{A}$ generates shifts of the temporal Gaussian wavepacket
$\exp\left[ -\I \tilde{t} \op{A} \right]\ket{\phi(0)}=\ket{\phi(\tilde{t})}$,
and $\tilde{\tau}$ plays the role of an interaction strength. After passing through the birefringent material and projection onto $\ket{\psi_0}$, the (unnormalized) state is
\begin{equation} \label{eq:11}
\ket{\Psi_1} =  \ket{\psi_0} \bra{\psi_0}\exp\left[ -\I \frac{\tilde{\tau}}{2} \op{O} \otimes \op{A} \right] \ket{\psi_0} \ket{\phi(0)}.
\end{equation}
This can be generalized to find $\ket{\Psi_{\ell}}$ , which is the state after $\ell$ repetitions of the interaction $\op{U}(\tilde{\tau})$ and the projection $\ket{\psi_0} \bra{\psi_0}$.
%

%
%

We now assume that the coupling between the system and the apparatus is weak, which means that the wavepackets corresponding to the two polarizations largely overlap. Mathematically this is expressed as $\tau \ll \tau_{G}$, or $\tilde{\tau} =\tau/\tau_G \ll 1$.
Since the interaction is weak, we can expand $\op{U}(\tilde{\tau})$ to second order in $\tilde{\tau}$. It can be shown that the final state after $\ell$ Zeno stages is then \cite{Aharonov_1993, Chen_2023}
\begin{equation}\label{eq:5}
\ket{\Psi_\ell} = \ket{\psi_0} \left[1- \frac{1}{2} \left(\frac{\tilde{\tau}}{2}\right)^2\Delta O^2 \op{A}^2\right]^\ell 
\ket{\phi(\ell\tilde{\tau}\langle \op{O} \rangle/2)}.
\end{equation}
This equation indicates that the temporal state corresponds to a single Gaussian wavepacket of duration $\tau_G$ that has been shifted in time by $\ell\tilde{\tau}\langle \op{O} \rangle/2$. Thus, if we measure this delay, after having calibrated $\tilde{\tau}$, we can directly determine the expectation value of the polarization operator $\langle \op{O} \rangle$. Since the delay of the wavepacket increases with the number of Zeno stages, but the width stays the same, the resolution of the measurement of $\langle \op{O} \rangle$ also increases with the number of stages.

It may be surprising that the output is a single Gaussian wavepacket, not two separate wavepackets. The reason for this is that the interaction is weak. After each passage through the birefringent medium the wavepacket splits into two wavepackets with two different polarizations. However, since the wavepackets strongly overlap, the projection back onto the initial polarization state $\ket{\psi_0}$ forces them to interfere with each other. The result is that after each Zeno stage we are left with a single Gaussian wavepacket whose width does not change appreciably, but whose temporal shift depends on the polarization. For example, if the state is vertically polarized the wavepacket will come out sooner, while if the state is horizontally polarized the wavepacket will come out later. If the state has equal horizontal and vertical components the wavepacket will come out half-way between these two extremes. Every passage through the medium amplifies the delay.

\subsection{Polarization stabilization}
To actively stabilize the polarization we use a stochastic parallel gradient descent (SPGD) algorithm \cite{su_2017}. The goal is to maximize the function $f(\mathbf{V})$, where $f$ is the measured photon count rate and $\mathbf{V}=(V_1,V_2,V_3,V_4)$ are the voltages applied to four fiber squeezers in a polarization controller. The basic idea is to make small, random changes $\delta \mathbf{V}$ to $\mathbf{V}$ while measuring $f$, in order to estimate the gradient of $f$ along the direction of $\delta \mathbf{V}$. $\mathbf{V}$ is then updated to move toward the maximum of $f$ and the process is repeated. The algorithm we use is given in Table~\ref{tab:SPGD}.

\begin{table*}
    \centering
    \begin{tabular}{p{0.06\linewidth} | p{0.7\linewidth}}
     Step & Action\\
    \hline
    1 & For a constant $C$, randomly assign $\delta V_j$ to be $C$ or $-C$ for $j = 1, 2, 3, 4$. Let
$\delta \mathbf{V}=(\delta V_1,\delta V_2,\delta V_3,\delta V_4)$.\\
    2 & Measure $f^+ = f(\mathbf{V} + \delta \mathbf{V})$ and $f^- = f(\mathbf{V} - \delta \mathbf{V})$, and let $\delta f = f^+ -f^-$.\\
    3 &  For a positive constant $\gamma$, define $G=\gamma \delta f$.\\
   4 &  For a positive constant $G_{\text{max}}$, if $G > G_{\text{max}}$ set $G = G_{\text{max}}$, if $G < -G_{\text{max}}$ set $G = -G_{\text{max}}$.\\
   5 &  Update  $\mathbf{V}$ to $\mathbf{V} + G \delta \mathbf{V}$.\\
    6 & Return to Step 1 and repeat.\\
\end{tabular}
\caption{\label{tab:SPGD} The steps used in our SPGD algorithm. The constants $C$, $\gamma$ and $G_{\text{max}}$ are determined empirically by monitoring the count rate to make sure that it is stable.}
\end{table*}

%
This is the same algorithm that was used in Ref.~\cite{su_2017}, except for the addition of step 4. We added this step because we found that fluctuations in the count rates would occasionally lead to large changes in the drive voltages that would send us away from the maximum. Limiting the maximum voltage change between steps eliminated this problem.

It may be the case that accumulated phase drifts cause the voltage determined by the algorithm to exceed the maximum (minimum) voltage allowed by the polarization controller. In this case, we simply subtract (add) a voltage corresponding to a $2\pi$ phase shift, bringing the voltage back within range. We find that this "unwinds" the phase sufficiently well for future iterations of the algorithm to correct any residual errors.

This algorithm runs continuously during our experiments, using the same photocounts that are used in our protective measurements (but with an independently adjustable integration time). There is no need to use time division multiplexing to separate the stabilization from the signal.

\section{\label{sec:expt}Experiments}

\begin{figure}[ht!]
\centering\includegraphics[width=7cm]{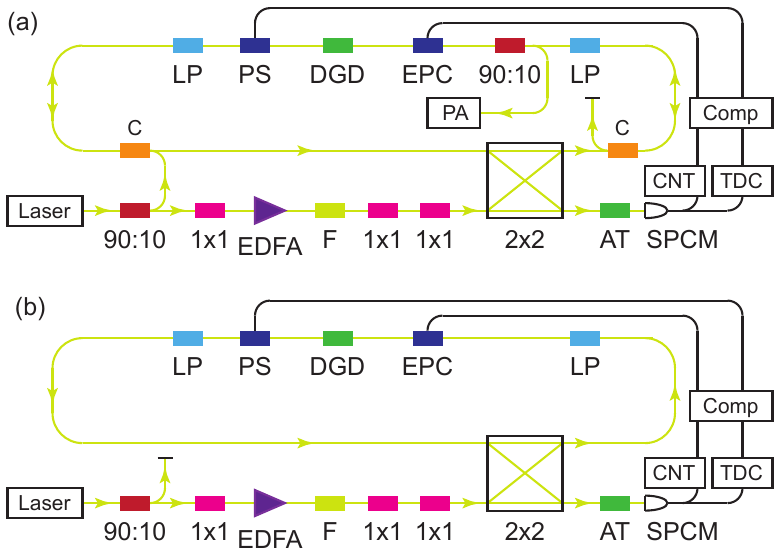}
\caption{\label{fig:Expt}The experimental apparatus. The abbreviations are: linear polarizer (LP), polarization stabilizer (PS), differential group delay (DGD), electronic polarization controller (EPC), 90\%/10\% splitter (90:10), polarization analyzer (PA), circulator (C), computer (Comp), counter (CNT), time-to-digital converter (TDC), amplitude modulator (1x1 switch), erbium-doped-fiber amplifier (EDFA), bandpass filter (F), 2x2 switch, variable attenuator (AT) and single-photon-counting module (SPCM). The arrangement in (a) is used to characterize the performance of the active polarization stabilization. The arrangement in (b) is used to perform protective measurements.}
\end{figure}

The experimental apparatus for characterizing the performance of our polarization stabilizer is shown in Fig.~\ref{fig:Expt}(a). This apparatus consists primarily of fiber-optical components connected by single-mode optical fibers. Light from a continuous-wave (CW) laser operating at \unit[1540]{nm} passes through a splitter, where 90\% of the light is directed to an amplitude modulator (a 1x1 switch). Three-nanosecond electrical pulses, at a repetition rate of between 34 and \unit[50]{kHz}, slice out optical pulses that are approximately \unit[2.5]{ns} in duration (full width at half maximum). These pulses are amplified by an erbium-doped-fiber amplifier, and a 100-GHz bandpass filter is used to reduce amplified spontaneous emission. The pulses then pass through two more amplitude modulators to further increase the contrast between the peak of the pulse and the CW background. A 2x2 switch transfers the pulses into a loop; by adjusting the timing of the switch, the pulses propagate a controllable number of times, counter-clockwise, around the loop before being switched back out. 

Inside the loop the pulses pass through a circulator, a linear polarizer, and then 90\% of the light passes through a splitter. A computer-controlled electronic polarization controller (EPC) allows us to control the polarization before the pulses enter the DGD ($\unit[\sim250]{m}$ of birefringent fiber). The EPC (Oz Optics, EPC 400) consists of four optical-fiber squeezers, and was chosen for its relatively low insertion loss when compared to electro-optical polarization controllers. After the DGD the pulses pass through the polarization stabilizer (PS), which is simply a second EPC that is controlled by the algorithm described above. The pulses then travel through a linear polarizer and a circulator before returning to the 2x2 switch, where they can be directed back around the loop or switched out. The pulse repetition rate is chosen to ensure that there is at most one pulse circulating around the loop at a given time (we use a lower repetition rate when making more passes around the loop). 

After the pulses are switched out of the loop they pass through a variable attenuator that ensures the average photon detection rate is $<0.1$ per pulse at a single-photon counting module (SPCM). This SPCM has a temporal resolution of \unit[100]{ps}. One output of the SPCM is directed to a simple counter that is used for the stabilization feedback. Another output is directed to a time-to-digital converter (TDC), which measures the photon arrival times with \unit[20]{ps} precision. 

The 10\% of the CW light that is split off from the laser is coupled into the loop via a circulator. This light travels clockwise around the loop, the opposite direction as the pulses. A splitter directs 10\% of this light to a polarization analyzer, which is used to monitor the state of polarization of the light passing through the PS, DGD and EPC. Finally, this CW light is coupled out of the loop, before it would get to the 2x2 switch, by a second circulator.

Figure~\ref{fig:Expt}(b) shows the configuration of the apparatus for performing protective measurements. The difference between the two configurations is that for protective measurements the circulators and the 90:10 splitter are removed from the loop, as they are only needed to characterize the stability of the polarization.
The linear polarizers, and the components between them in Fig.~\ref{fig:Expt}(b), are used to perform the Zeno PM. The losses in our loop are relatively high ($\unit[\sim7]{dB/loop}$) because some of the individual components have high losses. The losses are dominated by the 2x2 switch ($\unit[\sim2]{dB}$) and the linear polarizers ($\unit[\sim1.5]{dB}$ each). In order to overcome large losses after as many as 13 loops, we start with approximately $10^9$ photons per pulse. The attenuator before the SPCM ensures that we measure fields with $<0.1$ photons per pulse at our detector.

The purpose of the PS is twofold. First, it compensates for polarization drifts due to phase shifts as light passes through the DGD. Additionally, by maximizing the count rate, it also ensures that the initial polarization state $\ket{\psi_0}$ prepared by the combination of the first LP and the EPC is the state that is projected onto by the second LP. This is a necessary condition for a proper PM, as described in Sec.~\ref{sec:PMtheory}. That the PS accomplishes both of these is explained in detail in Ref.~\cite{Chen_2023}, but a relatively simple explanation is given by the following argument. The initial polarization state is $\ket{\psi_0}$. The PS essentially determines the state that is projected onto at the end of the Zeno stage. Maximizing the counts means that this projection is $\ket{\psi_0}\bra{\psi_0}$, which is exactly what we want for a successful PM. 

\subsection{Polarization stabilization}

\begin{figure}[ht!]
\centering\includegraphics[width=7cm]{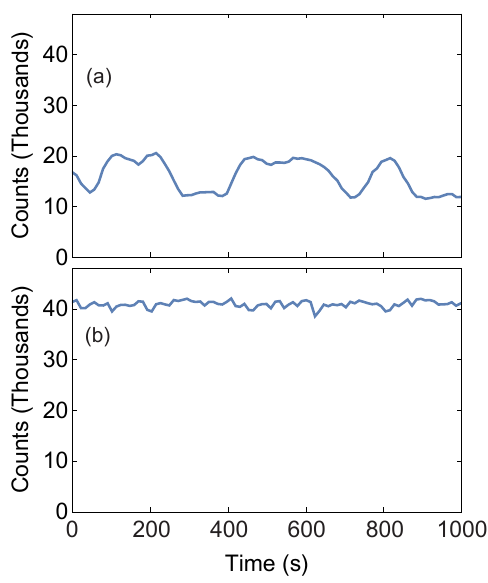}
\caption{\label{fig:Stable}The number of photon counts in \unit[10]{s} recorded when the stabilization algorithm is (a) off, and (b) on. The timing of the 2x2 switch was set to have photons propagate 5 times around the loop.}

\end{figure}

As shown in Ref.~\cite{Chen_2023}, a proper PM is performed by our apparatus when the measured photon count rate is maximized, so we use the count rate at the SPCM as the error signal for polarization stabilization. Figure~\ref{fig:Stable}(a) shows the measured photon counts as a function of time when the stabilization is turned off. It is seen that the count rate drifts on a timescale of 10's of seconds, due to phase shifts in the DGD that modify the polarization---polarization changes are converted to count rate changes by the second LP in the loop. When the stabilization is activated, the measured photon counts are shown in Fig.~\ref{fig:Stable}(b). It can be seen that the count rate is higher when stabilization is on, since the algorithm maximizes the count rate. It can also be seen that the fluctuations of the counts are significantly reduced. For the data in Fig.~\ref{fig:Stable}(b) the ratio of the standard deviation of the counts to the mean of the counts is 0.016. The integration time of the counter used in the stabilization loop was \unit[0.2]{s}, and the overall bandwidth of the stabilization loop is $\unit[\sim1]{Hz}$.

We have also demonstrated that stabilizing the count rate simultaneously stabilizes the polarization.  We have done this by measuring Stokes vectors describing the polarization using the polarization analyzer shown in Fig.~\ref{fig:Expt}(a). When the stabilization is off and the measured Stokes vector is plotted in the Poincar\'e sphere, we find that the polarization typically rotates in a circle on the sphere. It is this rotation that causes the intensity fluctuations shown in Fig.~\ref{fig:Stable}(a). When the stabilization is on we find that the measured Stokes vectors are clustered around a point on the Poincar\'e sphere. 

\begin{figure}[ht!]
\centering\includegraphics[width=7cm]{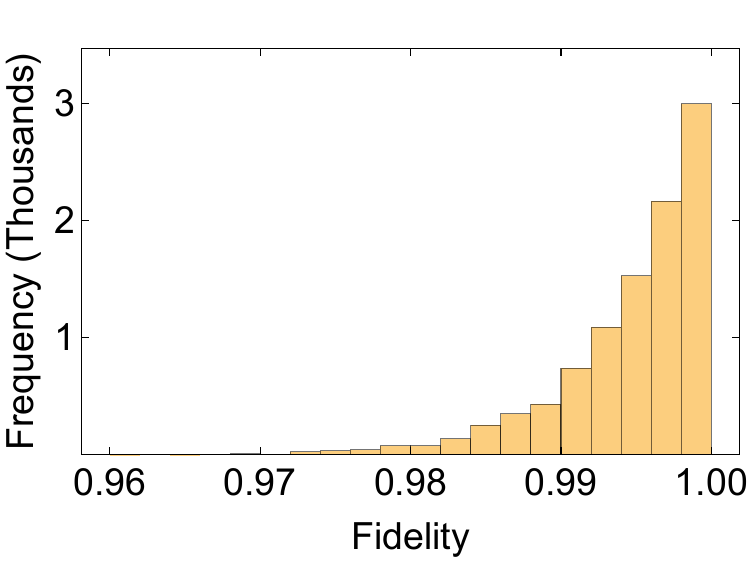}
\caption{\label{fig:Fidel}Histogram of the fidelity [Eq.~(\ref{eq:fidel})] between the mean Stokes vector and each individually measured Stokes vector. The data was acquired during the same time interval as that shown in Fig.~\ref{fig:Stable}(b).}
\end{figure}

By averaging the measured Stokes vectors, we can calculate the mean polarization. To determine the stability of the polarization about the mean, we use an expression motivated by the fidelity between two quantum states when expressed as vectors in the Bloch sphere. We define the fidelity between two polarizations as 

\begin{equation}
{F} = \frac{1}{2}(1+\mathbf{s}_1 \cdot \mathbf{s}_2)
   \label{eq:fidel}.
\end{equation}
Here $\mathbf{s}_1$ and $\mathbf{s}_2$ are the Stokes vectors corresponding to two different polarizations. Using measured Stokes vectors acquired during the same time interval as shown in Fig.~\ref{fig:Stable}(b), we compute the fidelity between the mean Stokes vector and each individually measured Stokes vector by applying Eq.~(\ref{eq:fidel}). A histogram of these fidelities is plotted in Fig.~\ref{fig:Fidel}. The mean fidelity for this data is 0.995, which corresponds to an angle between $\mathbf{s}_1$ and $\mathbf{s}_2$ of \unit[0.14]{rad}. The histogram indicates that 85\% of the fidelities are above 0.99 and 98\% of the fidelities are above 0.98. This demonstrates that our algorithm works very well at stabilizing the polarization.

\subsection{Protective measurements}

Having been satisfied that our algorithm for maximizing the count rate also stabilizes the polarization, we can now remove the circulators and the 90:10 splitter from the loop, switching from the configuration of Fig.~\ref{fig:Expt}(a) to that of Fig.~\ref{fig:Expt}(b). Note that in this configuration where we are performing PMs, we are still stabilizing the polarization by maximizing the count rate.

\begin{figure}[ht!]
\centering\includegraphics[width=7cm]{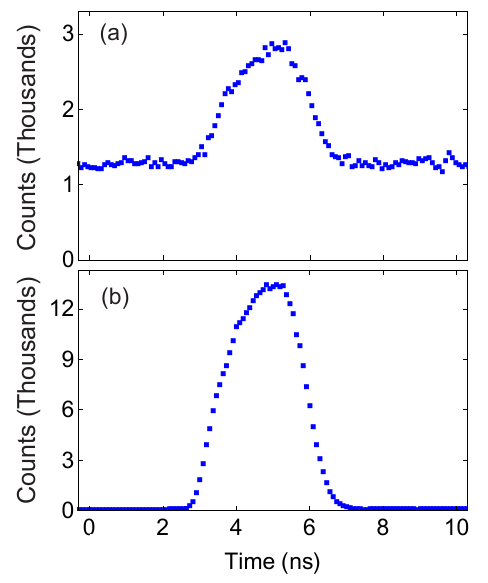}
\caption{\label{fig:Back}Measured photon arrival times without background subtraction or windowing. The data in (a) was acquired after 9 loops using a reference beam for polarization stabilization \cite{Chen_2023}, while the data in (b) was acquired after 13 loops using photon counts from the signal for stabilization [Fig.~\ref{fig:Expt}(b)].}
\end{figure}

As described above, to perform protective measurements we measure the arrival time of photons after a predetermined number of Zeno stages, with each stage corresponding to a loop in our experimental apparatus. Figure~\ref{fig:Back}(a) shows a histogram of measured arrival times after 9 loops using the experimental apparatus of Ref.~\cite{Chen_2023}, which employed a classical reference beam to stabilize the polarization. The large background in this plot is due to the classical reference, which limits the signal-to-noise (signal peak-to-background) ratio to $\sim2$. In this case the background limits the number of Zeno stages that can be used, because after 10 loops the signal has disappeared into the background. In Ref.~\cite{Chen_2023} it was necessary to subtract the background in order to determine the statistics of the arrival times of the signal photons.

Using our new apparatus [Fig.~\ref{fig:Expt}(b)] and polarization stabilization technique, a histogram of photon arrival times after 13 loops is shown in Fig.~\ref{fig:Back}(b). It can be seen that the background has been reduced to nearly 0 (signal-to-noise ratio $>100$), and this is one reason that we can perform measurements with up to 13 loops, as opposed to the maximum of 9 loops in our previous experiments \cite{Chen_2023}. With our new polarization stabilization technique, background no longer limits the number of loops in our experiments. We are instead limited by losses within the loop, as we need to detect a minimum of $\sim500$ counts per second for the fluctuations in the photon counting rate to be small enough for the stabilization algorithm to function properly. This minimum count rate is independent of the pulse repetition rate, within the range of rates that we use. Our new stabilization technique reduces the losses in our loop by eliminating the circulators; with the circulators we measure the loss to be $\unit[\sim10]{dB/loop}$, whereas without the circulators it is $\unit[\sim7]{dB/loop}$. This decreased loss also contributes to increasing the number of loops that we can use from 9 to 13. We expect that this increased number of loops will lead to increased performance for protective measurements, and below we demonstrate that this is indeed the case.

Another advantage of eliminating the reference beam is that we are able to acquire data much more quickly. When using the reference beam with large numbers of loops, the signal was small enough that a majority of the measured counts were background, so long averaging times were necessary. For example, the data in Fig.~\ref{fig:Back}(a) used \unit[600]{s} of averaging time, and we would sometimes average for as long as \unit[9000]{s} to acquire good statistics. By contrast, in Fig.~\ref{fig:Back}(b) we see that we have even more signal counts in only \unit[150]{s} when the background is eliminated. In the measurements described below we use \unit[150]{s} averaging times. 

In our new technique we do not subtract background, but we do window the data to remove data in arrival-time bins that are away from the peak of the histogram and contain only background, in in a similar manner to what we did in our previous experiments with a classical reference \cite{Chen_2023} (note that the data in Fig.~\ref{fig:Back} has not been windowed). We do this for two reasons. One is that even small numbers of background counts sufficiently far from the peak of the histogram can affect calculating moments (especially higher-order moments) of the distribution. A second is that there will $always$ be some windowing, if only because of the finite size of the data. Given this, it is important to be consistent in how the windowing is done. 
Starting from the
histogram peak, we move to shorter (longer) times, and
locate the first time at which the histogram is less than 0.5\% of the peak. We truncate the histogram at this point, removing all data corresponding to shorter (longer) arrival times. The histogram can now be normalized to obtain estimates of the probability $P_i$ for a photon to arrive in time bin $t_i$. From the probability distribution we can calculate the mean arrival time $t_\text{M}$ and its standard deviation.

Recall that for a PM to be successful, the interaction on each loop must be weak, which in our experiments means that the DGD per loop must be less than the pulse duration. In our experiments $\tau_{\text{loop}}=\unit[(0.483\pm 0.001)]{ns}$ is the measured DGD per loop \cite{Chen_2023}, which is less than 1/5 of the $\unit[(2.5\pm 0.1)]{ns}$ full-width-at-half-maximum duration of the pulse in Fig.~\ref{fig:Back}(b). An illustration of the overlap of the $H$- and $V$- polarized pulses after a single passage through the DGD is available in Fig.~4(a) of Ref.\cite{Chen_2023}.

\begin{figure}[ht!]
\centering\includegraphics[width=7cm]{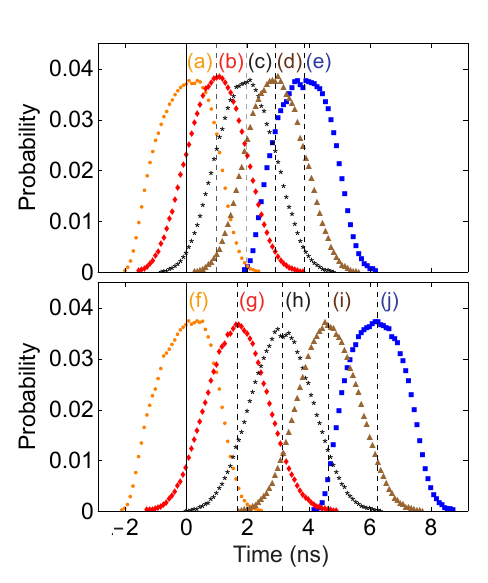}
\caption{\label{fig:Loops}Probabilities of photon arrival times after 8 loops (a)-(e) and 13 loops (f)-(j). Each curve corresponds to a different polarization. Orange circles (a), (f) correspond to polarization along the fast (vertical) axis of the DGD, blue squares (e), (j) correspond to polarization along the slow (horizontal) axis, and other curves are polarizations along intermediate angles. The vertical lines correspond to the mean arrival time for each polarization (given in Table~\ref{tab:values}).}
%
\end{figure}

In Fig.~\ref{fig:Loops} we show the probabilities of photon arrival times, using our new stabilization technique, after 8 loops and 13 loops. For 8 loops we use a pulse repetition rate of 50kHz, while for 13 loops we use a rate of 34 kHz. An arrival time of 0 is defined to be the mean arrival time for $\ket{V}$ photons. The $\ket{V}$ polarization state is aligned with the fast axis of the DGD, so as we rotate the polarization state from $\ket{V}$ to $\ket{H}$ the arrival time increases. However, as described above, the distribution of arrival times is determined by the duration of the input wave packet, and it stays nominally the same as the polarization changes. From the distributions shown in Fig.~\ref{fig:Loops} we calculate $t_\text{M}$, and these times are shown as vertical lines in the figure. It is seen that for a given polarization state, $t_\text{M}$ is larger for 13 loops than for 8 loops. From the theoretical discussion in Sec.~\ref{sec:PMtheory} we expect that $t_\text{M}$ will grow linearly with the number of loops, while the width stays approximately constant, and this is indeed the case. 

In order to convert measured mean time delays $t_\text{M}$ to expectation values $\langle \op{O} \rangle$ of the protective measurement, we scale the delay to the interval $[-1,1]$: $\langle \op{O} \rangle = 2\frac{t_\text{M}}{\tau_\text{max}}-1$. Here $\tau_\text{max} = \ell \tau_{\text{loop}}$, where $\ell$ is the number of loops. With this scaling we can use the arrival time probability distributions to calculate the mean of $\langle \op{O} \rangle$, as well as the uncertainty $\sigma_{\text{PM}}$ (as measured by the standard deviation) of the protective measurement.

In Table~\ref{tab:values} we give statistics for the measured arrival times and expectation values. The top section of this table gives results that we had previously obtained using 8 loops and a classical reference beam for polarization stabilization \cite{Chen_2023}. The bottom two sections correspond to the 8- and 13-loop data shown in Fig.~\ref{fig:Loops}. The first thing to note is that the standard deviations of the arrival times for the 8-loop data, whether stabilized using a classical reference or signal photons, are nearly the same. These values are on the order of 10's of ps smaller for stabilization using the classical reference. The corresponding uncertainties of the PMs $\sigma_{\text{PM}}$ are also nearly the same. The classical reference yields uncertainties that are smaller by about 0.01, indicating that it does a slightly better job of stabilizing the polarization. However, the results from the two stabilization techniques are remarkably similar, which is another indication that our new technique works well.
\begin{table*}
    \begin{tabular}{l|lcrccc}
     Stabilization (loops)  &   Distribution & $t_\text{M}$ (ns) & $\langle \op{O} \rangle$ & $\sigma_\text{PM}$ & $\sigma_\text{SM}$ & $\mathcal{R}$\\
    \hline
     & & $0.00 \pm 0.79$ & $-1.00$ & 0.41 & 0 & 0\\
     &  & $1.07 \pm 0.87$ & $-0.45$ & 0.45 & 0.89 & 2.0\\
    Classical reference (8) & see Ref.~\cite{Chen_2023} & $2.07 \pm 0.89$ & $0.07$ & 0.46 & 1.00 & 2.2\\
     &   & $3.04 \pm 0.85$ & $0.57$ & 0.44 & 0.82 & 1.9\\
     &  & $3.87 \pm 0.81$ & $1.00$ & 0.42 & 0 & 0\\
     \hline
     & (a) orange circles & $0.00 \pm 0.82$ & $-1.00$ & 0.42 & 0 & 0\\
     & (b) red diamonds & $0.99\pm 0.89$ & $-0.49$ & 0.46 & 0.87 & 1.9\\
     Photons (8) & (c) black asterisks & $1.95 \pm 0.92$ & $0.01$ & 0.47 & 1.00 & 2.1\\
     &  (d) brown triangles & $2.90 \pm 0.88$ & $0.50$ & 0.46 & 0.87 & 1.9\\
     & (e) blue squares & $3.87 \pm 0.81$ & $1.00$ & 0.42 & 0 & 0\\
        \hline
     & (f) orange circles & $0.00 \pm 0.84$ & $-1.00$ & 0.27 & 0 & 0\\
     & (g) red diamonds & $1.67 \pm 0.97$ & $-0.47$ & 0.31 & 0.88 & 2.8\\
     Photons (13) & (h) black asterisks & $3.13 \pm 1.00$ & $0.00$ & 0.32 & 1.00 & 3.1\\
     &  (i) brown triangles & $4.65 \pm 0.95$ & $0.48$ & 0.30 & 0.88 & 2.9\\
     & (j) blue squares & $6.25 \pm 0.83$ & $0.99$ & 0.26 & 0.14 & 0.5
\end{tabular}
\caption{\label{tab:values} Data for the arrival time distributions (a)--(j) shown in Fig.~\ref{fig:Loops}. The table shows: (i) measured mean delays $t_\text{M}$ relative to the zero defined by the mean arrival time for photons polarized along the fast axis, and with uncertainties given by the standard deviation; (ii) corresponding expectation values $\langle \op{O} \rangle$ of linear polarization; (iii) uncertainties $\sigma_\text{PM}$ for the PM obtained from the uncertainties in the arrival times, rescaled to the range $[-1,1]$ of the expectation value; (iv) uncertainties $\sigma_\text{SM}$ for the strong measurement are obtained from Eq.~\eqref{eq:sigma_SM}, using the corresponding expectation value $\langle \op{O} \rangle$; (v) relative measurement performance $\mathcal{R}$, see Eq.~\eqref{eq:R}.}
\end{table*}

Comparing the results for 8 loops versus those for 13 loops in Table~\ref{tab:values}, it is seen that the uncertainties $\sigma_{\text{PM}}$ are significantly lower for 13 loops than for 8. This shows that as we increase the number of loops the PM performs better, which is what we would expect. Switching from our previous stabilization technique to our new technique is what allowed us to increase the number of loops and obtain these improved results. 

We would like to compare the uncertainty, expressed as the standard deviation, of our PM to the corresponding uncertainty of a strong measurement (SM). A SM of polarization that corresponds to our PM is described by a measurement of the observable $\op{O} = \ketbra{H}{H} - \ketbra{V}{V}$. Such a measurement is typically performed by splitting the incident signal photons on a polarizing beamsplitter and counting the number of photons emitted at the H and V outputs. Since we do not actually perform an SM, we instead calculate the theoretically expected uncertainty $\sigma_{\text{SM}}$. This uncertainty is given by the standard deviation
\begin{equation}\label{eq:sigma_SM}
    \sigma_\text{SM}= \sqrt{\langle \op{O}^2\rangle - \langle \op{O} \rangle^2}  =\sqrt{1 -\langle \op{O} \rangle^2},
\end{equation}
since $\op{O}^2=\op{1}$ is the identity operator.
Assuming the same number of detected photons for both the protective and strong measurements, we define the relative measurement performance $\mathcal{R}$ between the SM and the PM as \cite{Chen_2023}
\begin{equation}\label{eq:R}
\mathcal{R} =  \frac{\sigma_\text{SM}}{\sigma_\text{PM}}.
\end{equation}
If $\mathcal{R}  >1$, then the PM provides a lower-uncertainty estimate of the expectation value than the SM. 

Table~\ref{tab:values} shows $\sigma_{\text{SM}}$ and $\mathcal{R}$, using the measured expectation value from the PM for $\langle \op{O} \rangle$ in Eq.~\ref{eq:sigma_SM}. It is seen that the uncertainty of the PM is smaller than that for the SM, while correspondingly $\mathcal{R} >1$, for polarization states that are not too close to the eigenstates $\ket{H}$ and $\ket{V}$ of the observable $\op{O}$. In this sense, the PM outperforms the SM for nearly all polarization states. In the most extreme case illustrated in Table~\ref{tab:values}, the uncertainty of the PM is three times smaller than that of the SM. Our measurements thus confirm other theoretical and experimental results on this subject \cite{Piacentini_2017, Chen_2023, Schlosshauer_2024}. We compare SMs and PMs using the same number of detected, rather than initial, photons because the losses in our experiments are high. However, high losses are an in-practice, not an in-principle problem. Theoretical results show that the performance advantage of PMs can be maintained, even when considering the same number of initial photons \cite{Schlosshauer_2024}. 

Additionally, there currently exists new technology, based on photonic integrated circuits (PICs) developed for quantum computing, that could dramatically reduce the experimental losses. For example, our 2x2 switch has a loss of $\unit[\sim2]{dB}$ and our linear polarizers have losses of $\unit[\sim1.5]{dB}$. PIC switches and polarizers having losses of order $\unit[0.1]{dB}$ have both been demonstrated \cite{alexander_2025, shahwar_2024}. Protective measurements using this technology would have significantly improved performance, and would also enable one to launch true single photons into a device that performs Zeno protective measurements with a temporal pointer.
 
\section{\label{sec:conclusion}Conclusions}

We have demonstrated a technique for stabilizing the polarization of signal fields in an optical fiber by using the detected signal photon counts themselves as the error signal in a feedback loop. Compared to a stabilization technique that uses a classical reference beam, this technique offers increased signal strength and significantly decreased background. These improvements allow us to decrease the averaging time in our experiments. They also allow us to increase the number of Zeno stages in a protective measurement from 9 to 13. This increase in the number of stages leads to a corresponding decrease in the uncertainty of the measurements. The number of Zeno stages that we can use is limited by experimental losses, but new technology technology exists that could dramatically reduce these losses  \cite{alexander_2025, shahwar_2024}, and further improve the performance of protective measurements.



\begin{backmatter}
\bmsection{Funding}
US National Science Foundation 2109964

\bmsection{Acknowledgment}
We thank M. Schlosshauer for many helpful discussions. Additional funding for this project came from Reed College. EP and AC acknowledge support from the Gordon and Betty Moore Foundation. 
The research activities of MB were also supported in part by the U.S. National Science Foundation
through its employee IR/D program. 

\bmsection{Disclosures}
The authors declare no conflicts of interest.

\bmsection{Data Availability Statement}
Data underlying the results presented in this paper are not publicly available at this time but may be obtained from the authors upon reasonable request.

\end{backmatter}


\bibliography{References}

\end{document}